\documentclass[a4paper,USenglish,cleveref, autoref, thm-restate]{oasics-v2019}

\title{Documentation Generation as Information Visualization} 

\author{Will Crichton}{Stanford University}{wcrichto@cs.stanford.edu}{}{}

\authorrunning{W. Crichton} 

\Copyright{Will Crichton} 



\category{} 

\relatedversion{} 

\supplement{}

\nolinenumbers 

\hideOASIcs  

\EventEditors{John Q. Open and Joan R. Access}
\EventNoEds{2}
\EventLongTitle{42nd Conference on Very Important Topics (CVIT 2016)}
\EventShortTitle{CVIT 2016}
\EventAcronym{CVIT}
\EventYear{2016}
\EventDate{December 24--27, 2016}
\EventLocation{Little Whinging, United Kingdom}
\EventLogo{}
\SeriesVolume{42}
\ArticleNo{23}

\begin{document}

\maketitle

\begin{abstract}
Automatic documentation generation tools, or auto docs, are widely used to visualize information about APIs. However, each auto doc tool comes with its own unique representation of API information. In this paper, I use an information visualization analysis of auto docs to generate potential design principles for improving their usability. Developers use auto docs as a reference by looking up relevant API primitives given partial information, or leads, about its name, type, or behavior. I discuss how auto docs can better support searching and scanning on these leads, e.g. by providing more information-dense visualizations of method signatures.
\end{abstract}

\section{Introduction}

\begin{figure}[hbtp]
    \centering
    \includegraphics[width=0.9\textwidth]{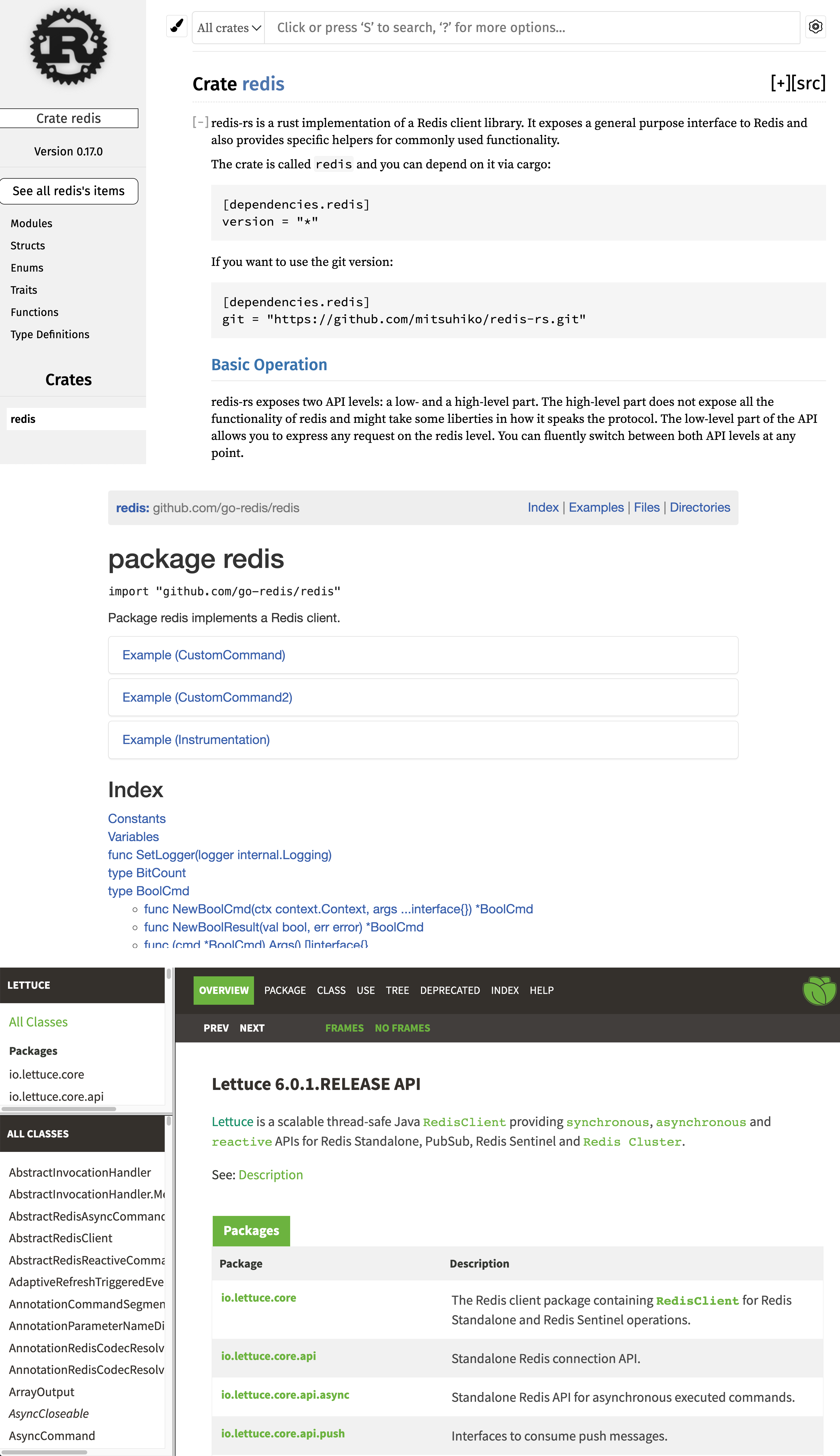}
    \caption{A screenshot of the front page of automatically-generated documentation for a Redis library in different languages. From top to bottom: Rust, Go, Java.}
    \label{fig:montage}
\end{figure}

Understanding other peoples' code is a fact of life in modern software development. With the rise of centralized package repositories, it's easier than ever to access third party libraries and frameworks. In 2019, the average JavaScript package had five direct dependencies and 86 transitive dependencies\,\cite{vaidya2019security}. Developers cite the availability of libraries as the \#1 reason to adopt a programming language\,\cite{meyerovich2013empirical}. Hence, making libraries easier to understand is a chief usability concern in today's programming landscape.

For the vast majority of libraries, their clients will never actually read the library source code. Instead, programmers rely on external resources: documentation, examples, tutorials, StackOverflow, blog posts, coworkers, and so on. However, these resources all have the same problem: they require a human. Someone has to write the blog, answer the StackOverflow post, and carefully craft the doc comments. Worse, that person probably has no training in technical communication, information visualization, computing education, or any discipline relevant to effectively explaining software. And most problematically, these resources can become out of date as soon as the library changes.

Automatically generated documentation, or auto docs, solve these problems by being derived directly from the source code. Tools such as Javadoc or Rustdoc take a codebase as input, and produce a wiki-like website as output. Except for doc comments, auto docs are always up-to-date. The author of the auto doc software can (potentially) present information according to best practices. And all this comes for free --- no extra effort needed by the library author! Every major programming language has a documentation generator, and recent languages consider one so vital that they ship it with the official toolset. Auto docs are arguably the most successful and widely-used software visualization tool in history.

Despite their apparent importance, auto docs are rarely the subject of research. They have few established best practices. Each new programming language comes with an entirely new docs format, as shown in Figure~\ref{fig:montage}. Beyond aesthetics, each interface provides significantly different means of visualizing and searching information. This variation raises the question: is one format better than another? In this paper, I show how an information visualization analysis can generate potential design principles for auto docs.

\section{Task Analysis}

To analyze auto docs from an information visualization perspective, we need to understand three things: the information, the tasks, and the representations that bridge the two. Then we can identify the gap between theory and reality to generate design ideas for future auto doc tools.
 
\subsection{What information is being represented?}

Auto docs generally present information about APIs, i.e. aspects that are externally visible to consumers of the library. APIs generally include:

\begin{itemize}
    \item \textbf{Data structures:} structs, enums, classes
    \item \textbf{Functions:} constructors, methods, standalone functions
    \item \textbf{Interfaces:} abstract classes, traits, typeclasses
    \item \textbf{Hierarchies:} modules, files, namespaces
\end{itemize}

These code constructs collectively describe the primitives of an API. A developer can usually provide source-code comments further explaining their purpose, which may contain prose description or code examples. Additionally, many relations arise from the structure of these objects. For example:

\begin{itemize}
    \item \textbf{Inputs:} a type is used as input to a function
    \item \textbf{Outputs:} a type is used as output of a function
    \item \textbf{Contains:} a type is a field of another type
    \item \textbf{Inherits:} a class inherits from another class
    \item \textbf{Implements:} a type implements an interface
\end{itemize}

Besides the explicit hierarchies within the language (e.g. namespaces), relations implicitly define hierarchies amongst API primitives. For example, the C++ auto docs tool Doxygen will generate an inheritance tree diagram of all classes in a codebase. Rust's auto docs will also show all the types that implement a trait on the trait's generated page.
 
\subsection{What tasks use this information?}

Auto docs work best as a reference. That is to say, auto docs provide detailed information about individual constructs within an API. By contrast, tutorials are more effective at showing how multiple API constructs can be combined to accomplish a larger task. Therefore we will focus on tasks that developers have specifically for reference information.

The key feature of a reference is its ability to guide a reader with partial information about their object of interest. For example, if a person remembers the first letter of a particular word in a textbook, they can use a glossary to find the relevant word and corresponding pages. More generally, a person starts with a lead (e.g. a letter), and a reference provides some level of guidance in converting the lead into more information (e.g. a table of words grouped by first letter). For auto docs, the question is then: what leads do developers use when seeking information? Here are some example scenarios:

\begin{itemize}
    \item A developer wants to turn a list into a set. Their lead is \textit{functions that return the Set type and possibly take the List type as input.}
    \item A developer wants to find an element of a list that matches a predicate. Their lead is \textit{the natural language keywords ``list'', ``find'', and ``predicate''}.
    \item A developer wants to know what immutable operations exist for lists. Their lead is \textit{methods on the List type that have a read-only modifier.}
\end{itemize}

Information foraging analyses have shown that programmers use leads to search documentation\,\cite{ko2006exploratory}. However, no prior work has compiled a holistic taxonomy of such leads --- an appropriate starting point for future work on auto docs.

\subsection{What is the best information representation for each task?}

Consider a developer with a lead of natural language keywords, e.g. the ``find'' and ``predicate'' example. They might adopt a range of strategies to forage for this information, for example in this order:

\begin{enumerate}
    \item \textbf{Search engine:} they go to Google and type ``<language> list find predicate''. No relevant results show up. 
    \item \textbf{Browser search:} they go to the auto doc page for the List data structure. They Ctrl+F in the browser for each keyword, but don't find a relevant method.
    \item \textbf{Scanning:} on the auto doc page, they scan through the list of method names and description. Realizing the method was called ``indexOf'' instead of ``find'', they locate the appropriate method.
\end{enumerate}

These examples highlights two key needs of a developer following a lead. First, they need to be able to encode the lead into a search. Second, if the search fails, they need a visualization of many possibly relevant objects which can be manually searched. Then the developer can make fuzzier associations between their lead and the provided information, e.g. the semantic relationship between ``indexOf'' and ``find''.
 
\subsection{Where do current information representations fall short?}

\begin{figure}[t!]
    \centering
    \includegraphics[width=\textwidth]{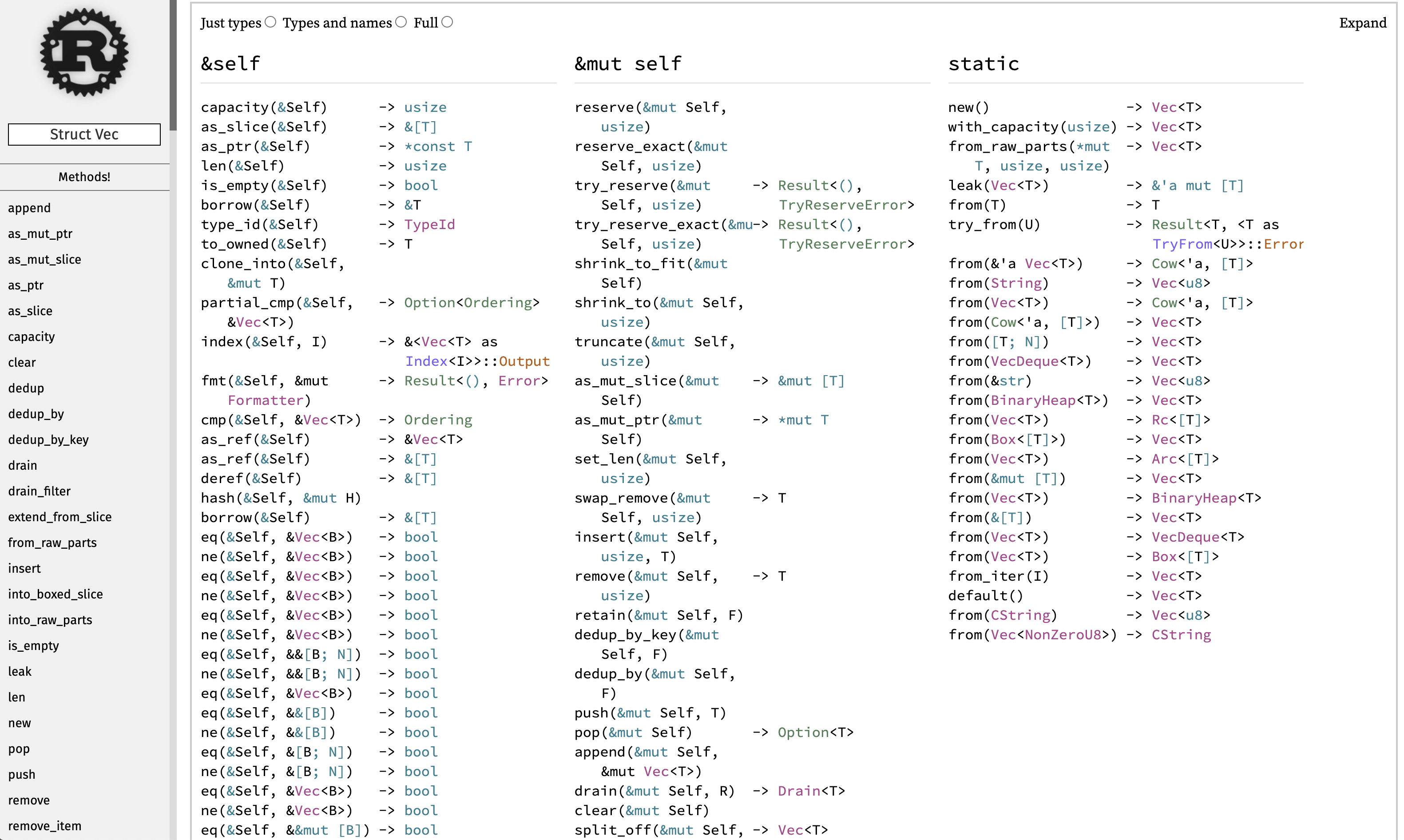}
    \caption{Prototype of a scanning-oriented interface for Rust's auto docs. By displaying methods in a table and grouping by the first argument, developers can quickly search for relevant methods.}
    \label{fig:prototype}
\end{figure}

For search, most auto docs tools support (at best) natural language keywords. But many relational leads are difficult to encode in a query. ``Functions that take a list as input and return an integer'' has a formal representation: the data type \verb|[a] -> int| where \verb|[a]| means a list of any kind of element. Notably, Haskell has a type-based search engine, Hoogle\,\cite{hoogle}, which enables these kinds of queries. But no other language or auto doc tool supports these queries.

For scanning, auto docs tools have two issues: first, scanning requires an initial filter or anchor to limit the set of possible matches. For most tools today, functions are anchored around their class. That means a developer can easily get a webpage of all the methods on the List class, but as mentioned above, they cannot easily get a page of all methods that return a List. Even within a class's method list, it can be valuable to filter. For example, Rust's documentation for \verb|Vec<T>|\,\cite{rustvec} lists well over 100 methods. The docs provides no way to e.g. filter for read-only methods that take \verb|&self| as input.

Second, auto doc tools do not necessarily provide efficient information representations to facilitate rapid scanning. Most auto docs today present methods in a one-dimensional list. However, a developer may more easily scan a large number of method names if they can all be on screen at once, e.g. in a two-dimensional table like the prototype in Figure~\ref{fig:prototype}. Alternatively, organizing methods hierarchically could turn a linear search into logarithmic. For example, the methods of a stateful class could be grouped around which states they apply to, which Sunshine et al.\,\cite{sunshine2014structuring} demonstrated will empirically reduce documentation search times for certain tasks.

\section{Conclusion}

My goal in this paper is to provide an initial framework for generating ideas about improving auto docs. Like many programming tools, auto docs are widely used, but have yet to be examined critically from a human-centered perspective. As Bret Victor points out in ``Learnable Programming''\,\cite{learnableprogramming}, an API resource should ``dump the parts bucket onto the floor.'' However, developers shouldn't have to spend hours searching through a sea of parts. With a careful understanding of what developers are looking for and what leads they use, we can build better tools for searching and visualizing API information.
 
\bibliography{oasics-v2019-sample-article}

\end{document}